# Strain Induced Incommensurate Structures in Vicinity of Reconstructive Phase Transitions

A L Korzhenevskii[1] and V Dmitriev[2,3]

[1] Institute for Problems of Mechanical Engineering, RAS, St. Petersburg, Russia

[2] Southern Federal University, Rostov-on-Don, Russia

[3] SNBL at ESRF, Grenoble, France

E-mail: dmitriev@esrf.fr

**Abstract**

General conditions controlling the formation of incommensurate phases (IPs) in crystals undergoing reconstructive phase transitions (RPTs) are analyzed in the framework of a model-free phenomenological approach. A universal trend to stabilizing such intermediate phases in vicinity of RPTs stems from the fact that certain high-order improper Lifshitz invariants reduce at RPTs to ones bi-linearly coupling critical displacement gradients and strains or even to the proper Lifshitz invariant. The approach developed here introduces a universal mechanism for the formation both of premartensite IPs and complex structures with giant unit cells, as found in some elemental crystals at high pressure.

It is convenient to assign phase transitions (PTs) to two main groups: (i) those where one of the phase structures (low symmetry) contains only symmetry elements belonging to the wider symmetry group of the other, parent one; and (ii) both neighboring phases contain symmetry elements absent in the group of the other one. The two above classes are termed as (i) preserving or (ii) breaking "group-subgroup" (GS) relationships [1]. For the PTs belonging to the first group, a very efficient symmetry based phenomenological Landau theory has been elaborated (see textbooks [2,3]). By contrast, for RPTs from the second group, breaking GS correlations, until recently no unifying theoretical approach had been suggested. Consequently, experimental data on RPTs were typically discussed in view of particular properties of specific materials, but not emphasizing common type anomalies, and without identifying features general for RPTs. However, it does not mean that the latter is a rare phenomenon – their occurrence in crystalline materials is on the same level as Landau ones. Numerous materials which are important for practical applications undergo RPTs, and the martensitic transformation (MT) in metals and alloys is a typical example [4,5].

A unifying crystal-geometrical theory of RPTs was suggested in late 1980s [6-8]. It has shown that for specific periodic sublattice shifts $\Delta$ along certain symmetric directions, new symmetry elements arise which break the GS relation. Thus, significant shifts can bring the crystal lattice, via RPT, to a new, non-group-subgroup related phase. Displacive mechanisms allow, therefore, to describe RPT using the "transcendental" order-parameter (OP) $\eta$ defined as a *periodic* function of the critical atomic displacements $\Delta$, and, consequently, to obtain a periodic thermodynamic potential F[T,P,$\eta(\Delta)$]. More details can be found in Ref.[1]. The minimization of F[$\eta(\Delta)$] with respect to $\Delta$ yields two types of stable phases: (a) "Landau" phases ($\partial F[\eta(\Delta)]/\partial\eta=0$) which are GS related to the parent phase symmetry, and (b) "limit" or "non-Landau" phases ($\partial\eta/\partial\Delta=0$) which have no GS relation to the parent structure and coincide with the extremes of $\eta(\Delta)$.

On one hand, this approach [1] was successfully applied to RPTs in some elements and compounds (see, for example, [1,9-11]). The specific topology of their P-T phase diagrams, the stability domains of the both non-Landau reconstructive phases and Landau ones, and the corresponding transition anomalies were explained in the framework of few unified phenomenological models.

On the other hand, diffraction experiments inside the premartensitic regions in some alloys undergoing RPTs (for instance, NiTi [12,13] or NaAl [14,15]) revealed anomalous superstructure reflections in vicinity of Bragg peaks: their incommensurate **q**-vectors have no inversion center, and its arrangement depends on the number of the corresponding Brillouin zone (BZ). Thus, the diffraction pattern does not correspond to a conventional periodic incommensurate structure. The phenomenological approach was typically employed in order to reveal the origin of such satellites [16,17]. It accounted for specific features of certain compounds [18,19], but neglected those distinguishing RPTs and Landau type PTs. The phenomenological models typically discussed premartensitic nucleation on defects [15,20,21], or considered martensite plates with a fine twin structure going down to the atomic scale (so called "adaptive phases") [22].

A new, not fully understood phenomenon, was discovered during the last decade by single crystal synchrotron radiation diffraction under high pressure. In elemental crystals, within a wide pressure range, complex crystal structures containing hundreds atoms in a unit cell were observed. The BaIV structure which contains, following [23], 768 atoms per unit cell is a record to date. The general symmetry based approach [1-3], however, predicts no such phases.

In this Letter we extend the approach [1] by accounting for the elastic properties of the crystal lattice. It enables us to reveal a universal scenario for the onset of IPs, induced by the RPTs. We will show its general thermodynamic origin based on the example of the low-temperature MT in the alkali metals. We will also briefly address the nature of complex structures observed recently in elemental crystals under high pressure.

The crystal-geometrical approach to RPT [1] considers the latter as a *uniform* transformation of a parent crystal lattice to a distorted one; the corresponding atomic displacements are mapped as shifts of the dimensionless points. Evidently, the consideration is missing inter-atomic interactions which are responsible particularly for the general capability of the crystal lattice undergoing deformations. Another specific point is that geometrical pathways at RPTs require sublattice shifts comparable to inter-atomic distances. One concludes therefore that, by analogy to the classic problem of the theoretical crystal strength [24], any advanced approach to RPTs should account for possible appearance of *inhomogeneous* structures. A realistic thermodynamic model (i) should deal with a thermodynamic (TD) potential *density*; (ii) the latter should contain not only the primary OP periodic contributions but also contributions of the strain field, including those induced by the non-uniform OP distribution.

Our unified approach can be introduced, without losing generality, through the example of the martensitic RPT BCC-9R [*Im-3m*(Z=1) to *R-3m*(Z=3)] which occurs in Li ($T_M$~77K), Na ($T_M$~35K) and, presumably, K metals [25]. The primary symmetry breaking BCC-9R mechanism relates to the twelve-dimensional irreducible representation (IR) $\tau_3(\mathbf{k}_4)$, with $\mathbf{k}_4 = \frac{1}{3}\left(\frac{\pi}{a}, \frac{\pi}{a}, 0\right)$ [10]. One finds that the corresponding TD potential contains "mixed" $5^{th}$-degree invariants which couple the OP and space derivatives $\nabla_i$ (IR $F_{1u}$) to shear strain tensor components $u_{jk}$ (IR $F_{2g}$) or to the symmetry-adopted $\varepsilon_1=(u_{xx}-u_{yy})/\sqrt{2}$ and $\varepsilon_2=(u_{xx}+u_{yy}-2u_{zz})/\sqrt{6}$ (IR $E_g$):

$$I_1^c = u_{xy} \cdot \nabla_z [ \left(\eta_1^3 - 3\eta_1^2\eta_2 - 3\eta_1\eta_2^2 + \eta_2^3 + \eta_3^3 - 3\eta_3^2\eta_4 - 3\eta_3\eta_4^2 + \eta_4^3\right) + \\ + \left(\eta_5^3 - 3\eta_5^2\eta_6 - 3\eta_5\eta_6^2 + \eta_6^3 + \eta_7^3 - 3\eta_7^2\eta_8 - 3\eta_7\eta_8^2 + \eta_8^3\right) ] + ... \quad (1)$$

In order to demonstrate their generic structure, one of four invariants is displayed in (1), the others appear in an effective form below. Only the first characteristic terms are shown in (1), the others can be produced by permuting subscript indexes.

In the literature, the identification of RPTs so far mainly underlined the break of the GS relationship between symmetries of the parent and non-Landau phases. However, little attention has been paid to the less evident fact that the *purely geometrical* conditions exist, and they control, in the non-Landau phase, the values of the OP and spontaneous strain components so that their magnitudes are independent of external parameters (temperature, pressure etc.) [1].

Therefore, the effective degree of the invariants of type (1) in the TD potential lowers, and this latter dramatically increases the impact of high-order mixed gradient terms.

The symmetry change Im-3m to R-3m requires for the twelve-component OP ($\eta_i$, i=1÷12) satisfying in the 9R phase condition:

$$\eta_1=\eta_2=\eta_0 \neq 0, \quad \text{and} \quad \eta_j=0 \text{ for } j=3\div 12, \tag{2}$$

where the OP equilibrium value $\eta_0$=const. Taking into account that the spontaneous strain components $u_{xy}^0$ and $\varepsilon_2^0$ are also constant in 9R [10], one immediately realizes that the corresponding coupling invariants assume the form similar to the proper Lifshitz invariant:

$$I_{1c}^{eff} = u_{xy}^0 \cdot \eta_0^2 \cdot (\nabla_z\eta_1 + \nabla_z\eta_2), \quad I_{2c}^{eff} = \eta_0^2 \cdot \left[\sqrt{3}\cdot\varepsilon_1\cdot(\nabla_x\eta_1 - \nabla_x\eta_2) + \varepsilon_2^0\cdot(\nabla_y\eta_1 - \nabla_y\eta_2)\right]. \tag{3}$$

We stress that, in contrast to Landau type PTs, the above reduction does not lead to a necessary small size of the renormalized coupling constants generally. Then one can write down an *effective* TD density in the form ($\eta_1=\eta_2=\eta$):

$$F[\eta, u_{ik}] = F_1[\eta] + F_2(u_{ik}) + F_3[\nabla\eta] + F_4(\eta, u_{ik}\nabla\eta). \tag{4}$$

The $F_1[\eta(\Delta)] = a_1\eta^2 + a_2\eta^4 + a_3\eta^6$ term is a non-equilibrium contribution of the primary OP. The OP is a periodic function of critical displacements $\Delta$ of corresponding atomic layers along the [-110]$_C$ direction, e.g. $\eta(\Delta) = \eta_0 \cdot \sin(3\pi\Delta/a\sqrt{2})$ [10]. $F_2$ is the elastic strain energy; it includes important improper terms coupling the primary OP {$\eta_i$} to shear strain tensor components $I_s^{eff} \approx (\eta_1^2 + \eta_2^2) \cdot u_{xy}$, and to its diagonal components $I_e^{eff} \approx (\eta_1^2 + \eta_2^2) \cdot (\varepsilon_1 + \varepsilon_2)$. $F_3[\nabla\eta]$ accounts for the conventional OP gradient contribution. Finally, $F_4$, being proportional to $I_{jc}^{eff}$ (3), is a coupling energy of the non-uniform OP distribution and a strain field. It should be specially noticed that the proper Lifshitz gradient invariant does not appear in (1) due to the symmetry restrictions. Nevertheless, the above symmetry adopted terms linear in the OP gradient ensure appearance of the stable IP. The latter can be driven either by spontaneous strains, or by the deviatoric stresses due to the non-hydrostatic compression conditions.

Already the equilibrium conditions (2) and the form of the inhomogeneous energy contributions (3) allow us to predict some crystal-geometrical characteristics for the strain-induced IP. The equation (2) identifies the atomic plane (110)$_{BCC}$ as becoming a hexagonal close-packed (001)$_{9R}$, and directs new three-fold axe along [110]$_{BCC}$∥[001]$_{9R}$ [10]. The coupling terms in $F_2$ predict that the spontaneous strains $u_{xy}$ and $\varepsilon_{1-2}$ should be induced in a new structure. The former, according to $I_{1c}^{eff}$, may impose, in the rhombohedral phase, incommensurate structure modulation along the [001]$_{BCC}$ direction, while the latter (see $I_{2c}^{eff}$) does the same in the [110]$_{BCC}$∥[001]$_{9R}$ direction.

To identify the IP structure in detail, one has to solve not simply algebraic but rather the partial differential equations of state coming from the functional (1), with certain boundary conditions. Generally very difficult, the problem can be solved in some special cases. In particular, if the strain distribution is approximately uniform, a triple sine-Gordon equation should be solved.

Its simplest solutions describing one-dimensional IPs, read:

$$x = \int_{x_0}^{\Delta} \frac{du}{\sqrt{u(1-u^2)(E + a_1 u - a_2 u^2 - a_3 u^3)}}, \qquad (5)$$

where $x_0$ and E are the integration constants.

Possible IPs are given by inverting the hyperelliptic integral $x=x(\Delta)$. Generally, it yields several periodic solutions, each defined by the couple of adjacent real roots of the denominator zeroes in (5) keeping the positive radical. The period of each IP structure is given by the numerical value of (5), where the high and low integration limits are equated to the chosen roots. Minimizing the potential (4), one finds the relevant solution $\Delta_0(x)$ [26]. It depends on the phenomenological parameters $a_1$ and $E$, and the magnitude of the uniform strain component coupled to the OP gradient. The component plays a role of the chemical potential for solitons in IP, and it controls stability limits of the latter. Notice that onset and evolution of the strain induced IP at RPT shows a clear analogy to the formation of soliton superstructures at the Landau type PT when the proper Lifshitz invariant is allowed [3]. The only new mathematical feature is that for RPT the full set of periodic solutions contains few distinct branches corresponding to different root couples, the solutions are hyperelliptic integrals. These are rare in problems in physics, while elliptic ones are common (see, however, note [26]). Concerning the physics, the difference is much more remarkable as the existence of such branches reflects the possibility of PTs between different IPs. The analysis of such sequential PTs observed, for example, in Eu [28] could be an exciting topic itself; however, it is beyond the subject of the present communication.

Another scenario comes into play if, after elimination of inhomogeneous strains, the sign of the renormalized coefficient at the OP quadratic gradient invariant becomes negative. Then in the vicinity of a continuous transition to IP, the corresponding linearized equations of state assume the same form as those near the Lifshitz point at the Landau type PT. Nevertheless, the fully developed IP structure is different since the TD potential at RPT is more complex than the ones describing the conventional Landau's IPs.

It is worth comparing our approach to the already existing models of inhomogeneous intermediate states observed at RPTs. In the framework of traditional crystal-geometrical models of strongly discontinuous MTs, the origin of anomalous satellites was attributed to the stacking faults accommodating the distortions of twinned martensite plates to the austenite matrix [22].

Polynomial expansion by strains is often used as a TD potential for crystals undergoing MTs (see, for example, [29]). It corresponds actually to the case of proper ferroelastic PT of the Landau type. In this case inhomogeneous structures are considered to be induced by defects. Following a similar strategy, i.e. inventing an analog of improper ferroelastic PT, a model was used in [16,17] whereby a phonon mode, corresponding to atomic layer shuffling at MT, is a primary OP which is coupled to the secondary uniform strains. One notices that both approaches are missing essential features of RPT and their OP. The former approach fully neglects the contribution of the actual OP in the TD potential (4) and the free energy is minimized only with respect to the strain field components. Obviously, it allows one to determine the mesoscopic structure morphology in the phase coexistence region only if the contribution of the elastic energy is dominant in the potential. The latter approach suffers because of the crude approximation of the displacement field $\Delta(x)$ by a phonon in a BZ point with rational coordinates, defined by a commensurate structure of the martensite phase. This is irrelevant to the case, as generally the spatial dependence of $\Delta(x)$ minimizing periodic TD potential at RPT cannot be described as a "frozen" phonon. In particular, the fact has been confirmed for Li experimentally [30]. Moreover, the unit cell parameters of the Li lattice in austenite (Im-3m) and martensite (R-3m) phases allow a practically perfect coherent habit-plane interface between the parent BCC structure and an *unfaulted* 9R phase. It would be logical to conclude that a simple crystal-geometrical approach is an adequate tool, while MT could occur via sidewise growth of the perfect martensite plates. However, instead of the latter, a mosaic of irregular martensite segments arises showing the presence of highly faulted or disordered polytype structures. It should be mentioned that the authors of Ref.[30] used *high-purity* samples while in experiments with Li-based solid solutions no similar close packed polytypes were observed [31]. It appears to reveal the inherent character of the polytype formation mechanism, while impurity suppresses the process. Thus, it is natural to believe that it is the existence of the above improper Lifshitz invariants that cause the instability of 9R with respect to complex polytype structures.

It would be worthwhile also to revisit the results of neutron diffraction studies of single crystalline potassium. The corresponding patterns were obtained, analyzed in detail, and discussed in late 1980s (see, for example, Refs.[32-35]). On one hand, weak superstructure satellites of (0.995, 0.975, 0.015) type were suggested by the authors of [32] to indicate the beginning of a BCC-9R transformation. On the other hand, in [33-35], it was interpreted as charge density wave satellites. The contradiction was not unambiguously resolved, while the above conclusion on stability of the complex incommensurate structure induced by the secondary elastic OP might provide a new understanding of the experimental data. Indeed, the directions of incommensurate modulations concluded from minimization of the inhomogeneous energy contributions $I_{1c}$ and $I_{2c}$ correspond fairly well to those observed in neutron scattering experiments.

General symmetry requirements imposed to OP which imply the mixed coupling term $\{\eta_i^n u_{ik}\nabla\eta_j\}$ should be derived following standard group theoretical procedures (see, for example, Ref.[2]). Comparing to the symmetry restrictions on the Lifshitz invariants, selection rules for the new gradient terms are less restrictive. Indeed, the symmetry properties of the conventional Lifshitz invariants $\{\eta_i\nabla\eta_k\}$ are described by the tensorial product of the antisymmetrized square $\{\eta_i^2\}$ of the OP representation and of the vector representation spanned by the coordinates $x_i$. For such term to be invariant, the antisymmetrized square of the OP should contain the vector representation V: $\{\eta_i^2\}\supset V$. In the above case of RPTs, the gradient operator $\nabla$, spanning V, transforms to $u_{ik}\nabla$, and the corresponding representation is a direct product of the one, $[U_{ik}]$, spanned by the components of the symmetric second-rank tensor and V. The above condition transforms to $\{\eta_i^2\}\supset V\otimes[U_{ik}]$. The product, in the general case, is a *reducible* representation, and $\{\eta_i^n\}$ ($n\geq 2$) can contain any of its irreducible parts.

For example, in the cubic point group m-3m, IR $F_{1u}$ is the vector representation, and it has to be present in the expansion of $\{\eta_i^n\}$. However, the operator $u_{ik}\nabla$ transforms as $F_{2g}\otimes F_{1u}=A_{2u}\oplus E_u\oplus F_{1u}\oplus F_{2u}$, and any of four latter IRs, but not exclusively $F_{1u}$, can be in $\{\eta_i^n\}$. It is worth noting that the above condition formally forbids the existence of the coupling invariants but does not forbid totally the onset of an IP, since the latter can arise due to energetic but not symmetry reasons.

Generalizing, one can predict, for RPTs whose OPs allow high-order invariant $\{\eta_i^n u_{ik}\nabla\eta_j\}$, a tendency for IP to occur under the application of large stresses, or in a coexistence range of strain-dominant MT, or its onset in plastically distorted crystals. It is promoted by the increase, in above cases, of impact of the mixed terms responsible for the formation of strain induced inhomogeneous structures.

The widespread extremely complex structures recently discovered in elemental crystals under high pressure (see review papers [36,37]) seem to be a good example of the trend. In particular, analyzing the exciting case of the Ba metal, one finds, first of all, that it undergoes the classical RPT BCC-HCP (*Im-3m* – *P6$_3$/mmc*) at $P_{tr}$=5.5 GPa. Then, at $P_i$=12 GPa, it adopts a very complex structure containing hundreds atoms per unit cell. Moreover, the corresponding so-called BaIV phase actually has a sequence of several different complex inhomogeneous structures [23]. Applying the above selection rules to the OP for the BCC–HCP transformation, which spans IR $\tau_4(\mathbf{k}_9)$ of the Im-3m space group, one finds that its TD potential does contain the coupling terms of the type (3) but the proper Lifshitz invariant is not allowed. This example supports our conjecture that many high-pressure complex phases should be treated as IPs induced by the improper Lifshitz invariants. It is worthwhile mentioning that this idea is not totally new (a suggestion to consider the specific structure Ga-II, assumed to contain 104 atoms/unit cell [38] as an incommensurate structure was published recently [39]) but now it also finds strong general symmetry and TD arguments.

It is worth recalling that both the form of allowed gradient invariants and the structure of possible IPs is controlled by the symmetry of the stress tensor. Consequently, the distribution of diffraction superstructure satellites depends on the compression conditions: the arrangement of satellites obtained in hydrostatic conditions should be different from the one observed at any deviation from hydrostaticity, whatever induces it. In any case, more experimental and theoretical efforts are needed in order to shed light on the origin of IPs related to RPTs, in general, and to disclose the nature of the complex structures observed in elemental crystals under high pressure, in particular.

Summarizing, we emphasize that the approach developed in the paper is model free. It is symmetry based and addresses the fundamental properties of RPTs. Both the periodic dependence of OP $\eta(\Delta)$, and the existence of effective improper Lifshitz invariants in the corresponding TD potential, are controlled by the crystal symmetry. It does not depend on the material and is equally applicable to metals, semiconductors or insulators. Due to weak symmetry restrictions for mixed gradient invariants, relevant to RTPs, the strain induced IPs seem to be more probable than classical Landau's ones. The general symmetry arguments are aiming at encouraging the systematic construction of improper Lifshitz invariants and carrying out first-principal calculations of the coupling constants for crystals undergoing RPTs.

**Acknowledgements**

Prof V. Shirokov is warmly acknowledged for providing us access to his group-theoretical software package and kind assistance in computer calculations. We would like to thank Dr. P. Pattison for a critical reading of the manuscript. A.K. research was supported by the RFBR through project N 13-02-91332.

solid body rotating around a fixed axis. Few years ago the method was applied to describe a particle motion in the multidimensional gravity field (for example, near the "black hole") [27].